# A simple argument that small hydrogen may exist

**J. Va’vra**

SLAC (Stanford Linear Accelerator Center), Menlo Park, CA 94025-701, USA; jjv@slac.stanford.edu

**Abstract**

This paper examines whether a compact electron–proton configuration (“small hydrogen”) with a characteristic radius of a few femtometers is excluded by basic relativistic kinematics and simple stationarity constraints. Motivated by earlier discussions of formally deep relativistic energy scales in Dirac-based treatments, a phenomenological, virial-inspired energy-balance framework that incorporates relativistic kinetic energy, finite-size regularization of the central field, and order-of-magnitude spin–magnetic and spin–orbit contributions is developed in this paper. Within this framework, self-consistent characteristic scales associated is obtained with a hypothetical compact configuration without invoking Dirac or quantum-electrodynamics (QED) bound-state eigenvalues. The resulting scales—namely, a central energy scale of about 260 keV and a characteristic spin-dependent scale of order $\Delta E_{spin} \approx 100 \pm 20$ keV—define concrete experimental and observational energy ranges of interest. The present study does not establish the existence, formation probability, lifetime, or dynamical stability of such states. Rather, it shows that relativistic kinematics, finite-size effects, and virial-inspired stationarity constraints do not, by themselves, rule out compact stationary electron–proton configurations within the assumptions of the model. If such states were realized in nature and possessed radiative or interaction channels, those states may have implications for astrophysics, fusion concepts, and dark-matter phenomenology.

## 1. Introduction

In the early development of nuclear and quantum physics, the possibility of electron–proton (e–p) compact states involving electrons and protons was occasionally discussed. In 1920, Ernest Rutherford speculated that a neutral nuclear constituent might consist of a proton and an electron bound considerably short distances [1]. Following the experimental discovery of the neutron in 1932, this hypothesis was examined and subsequently abandoned on fundamental grounds related to spin, statistics, beta decay, and the uncertainty principle [2]. The neutron is now understood to be a composite hadron with no electron content. This historical episode is mentioned solely to illustrate that short-distance e–p dynamics were explored in early quantum theory; it does not imply physical continuity between compact e–p configurations and the neutron, whose hadronic nature is firmly established.

Within relativistic quantum mechanics, the Dirac equation for a point-like Coulomb potential admits formal mathematical solutions that do not correspond to physical eigenstates, because the associated wave functions cannot be normalized. As a result, the notion of compact electron–proton compact states were set aside in mainstream atomic physics [3].

Interest in this question re-emerged sporadically in the later study [4,5], which highlights formal relativistic energy scales but does not establish the existence of hypothetical bound states, motivated by the observation that at distances comparable to the proton charge radius, the Coulomb potential must be replaced by a finite-size nuclear potential, such as the Smith–Johnson or Nix forms commonly used in relativistic Hartree–Fock calculations of heavy atoms [6,7]. These studies highlighted the appearance of formally deep energy scales in relativistic treatments, sometimes referred to as "deep Dirac levels" (DDLs). However, such approaches did not establish the existence of physically acceptable hypothetical bound states, nor did they address stability in a fully self-consistent manner.

It is now commonly recognized that a single-particle Dirac equation is insufficient to describe e–p dynamics at femtometer scales. At such distances, a proper treatment requires a relativistic two-body formulation incorporating quantum electrodynamics and proton structure effects. Stanley Brodsky argued that the Dirac equation in single-particle form is inadequate at these distances, and instead advocated a two-body QED treatment using Bethe–Salpeter-type approaches [8]. John Spence and James Vary implemented a two-body relativistic framework and reported indications of a bound solution [9], but did not pursue a definitive conclusion, in part due to the technical complexity of the required calculations.

The complication of the problem is therefore twofold: (a) there is no experimentally confirmed signature of compact e–p hypothetical bound states, and (b) a complete two-body QED (quantum electrodynamics) plus QCD (quantum chromodynamics) calculation at femtometer scales remains technically challenging and model-dependent. The present paper, does not considers a full two-body QED, or QED plus QCD solution, nor does claim to replace or approximate such a treatment. Instead, the study addresses a logically prior and more limited question: whether basic relativistic kinematics, finite-size nuclear potentials, and stability requirements forbid the existence of a compact stationary e–p configuration. To this end, here, an approximate but physically constrained framework is employed based on

1. the relativistic virial-stability condition,
2. the de Broglie standing-wave condition for a relativistic electron, requiring that the orbital circumference accommodates an integer number of de Broglie wavelengths, $2\pi r = n \cdot \lambda_{\text{de Broglie}}$, and
3. the requirement of negative total binding energy.

This framework is not intended to establish the existence, formation probability, or lifetime of such a configuration, but rather to examine whether relativistic kinematics and stability constraints alone preclude it; it provides an internally consistent estimate of the characteristic radius, binding energy, and hyperfine scale of a hypothetical compact state, without invoking the Dirac spectrum as a source of physical eigenvalues.

A tightly bound e–p configuration cannot form spontaneously from the static Coulomb field alone, since the maximum electrostatic energy available at the radii of a few femtometers is about 0.5 MeV. Formation would therefore require an external energy input, analogous in scale to processes such as electron capture (p + $e^-$ → n + $\nu_e$), which requires an energy threshold of 0.782 MeV.

In Section 3, the conditions for relativistic virial stability of a compact e–p esystem are analysed and the resulting energy scales are derived. The role of the Dirac equation in Section 2 is discussed only as historical motivation; the quantitative results of this paper do not rely on Dirac eigenvalues or on the existence of physical deeply bound Dirac states.

Given the absence of a tractable first-principles two-body QED solution at femtometer scales, it is meaningful to raise the more limited question of whether relativistic kinematics and stability conditions alone forbid such configurations.

## 2. Dirac Equation Effort

Reference [5] applied the Dirac equation to explore whether tightly bound e–p statesmay exist. The Dirac spectrum for a Coulomb potential contains two mathematical branches:

$$s = s(\pm) = \pm\left(\left(j + \frac{1}{2}\right)^2 - \alpha^2\right)^{1/2} \quad (1)$$

with total energy levels according to Sommerfeld–Dirac approach [11–13]:

$$E = mc^2\left(1 + \frac{\alpha^2}{(s+n_r)^2}\right)^{-1/2}, \quad (2)$$

where $j = \ell + s$, $s = \pm 1/2$ is a spin, $\ell = k - 1$ is the orbital momentum, $m$ denotes the electron mass, $\alpha = e^2/\hbar c = 1/137$ represents a fine structure constant, $e$ denotes the elementary electric charge, $c$ is the speed of light, $\hbar$ is the reduced Planck constant, $n_r = 0,1,2,3,\ldots$, and $k = 1,2,3,\ldots$

Motivation from finite-size regularization of the Dirac–Coulomb problem is as follows. For a point-like Coulomb field, the Dirac equation admits the known physical branch that produces ordinary hydrogenic eigenstates, while the formally "deep" branch does not yield an acceptable bound-state solution. References [4,5] emphasize that this conclusion can be actually altered if the proton is treated as an extended charge distribution, which regularizes the short-distance behavior of the central field. The proposed strategy is to solve the Dirac equation inside a finite-radius proton model and outside in the Coulomb region, and to impose continuity of the Dirac radial spinor components at the matching radius. This program was suggested as a possible route to restoring mathematical acceptability of deep solutions, but it has not been successfully demonstrated in a definitive two-body bound-state calculation. In the present paper, these historical considerations serve only as motivation; no Dirac deep-branch eigenvalues are used, but instead a separate virial-inspired energy-balance consistency test is applied.

Attempts to obtain compact bound-state solutions directly from the single-particle Dirac equation with finite-size nuclear potentials did not yield stable, square-integrable states. This outcome is consistent with the general viewpoint emphasized by Brodsky, namely that relativistic e–p hypothetical bound states at short distances require a full two-body quantum-field treatment rather than a single-particle Dirac equation [10]. The appropriate framework is a two-body QED bound-state equation, such as the Bethe–Salpeter formulation [8].

John Spence and James Vary implemented a relativistic two-body QED framework and reported indications of a deeply bound solution [9]. However, the treatment assumed a point-like proton and did not incorporate proton structure or higher-order field-theoretic effects. Authors did not pursue the solution further due to its computational complexity.

The present study proceeds using a different approach. The remainder of this paper employs a relativistic virial-theorem framework incorporating relativistic kinematics, finite-size nuclear potentials, and stability conditions, without invoking Equations (1) and (2) for quantitative predictions.

Since a complete two-body quantum-field-theoretical treatment of the e–p system—including relativistic dynamics, proton structure, and radiative effects—remains mathematically complex and model-dependent, the present study addresses a different and logically prior question.

Specifically, the paper examines whether relativistic kinematics, finite-size nuclear potentials, and basic stability requirements forbid the existence of a compact stationary electron–proton configuration. This is done using a virial-theorem-motivated relativistic consistency framework, which provides a necessary (though not sufficient) diagnostic test, independent of detailed QED dynamics..

## 3. Simple Argument for Small Hydrogen

Let us examine whether a compact stationary e–p configuration is *kinematically and energetically forbidden* once relativistic motion, finite-size effects, and basic stability conditions are imposed. The analysis does not derive a hypothetical bound state from QED or from the Dirac equation, but instead applies a relativistic virial consistency condition to explore whether such configurations are excluded on general grounds.

Because an electron confined to femtometer scales is necessarily highly relativistic (with the Lorentz factor $\gamma_e \gtrsim 100$), nonrelativistic hydrogenic expressions are inapplicable. Let us therefore consider a phenomenological energy-balance framework in which the total interaction energy is expressed schematically as the sum of dominant contributions:

$$U = V_{\text{eff}}\,(r) + V_{(\text{Spin}.B)}(r) + V_{\text{SO}}(r),$$

where each term is evaluated only as an order-of-magnitude contribution at radii $r \approx 1$–$5$ fm, a regime where relativistic motion and finite-size effects dominate. The term $V_{(\text{Spin}.B)}(r)$ takes opposite sign for aligned and anti-aligned spin configurations.

### *3.1. Effective Coulomb Potential* $V_{\text{eff}}$

An entirely Coulomb interaction $V_C = -Ke^2/r$ (with $K$ a constant) does not yield a compact femtometer-scale solution in the present analysis.

The following expression earlier considered in exploratory treatments of compact e–p configurations [14,15],

$$V_{\text{eff}} = \gamma_e V_{\text{C}}(r) - \frac{V_{\text{C}}(r)^2}{2m_e c^2}, \tag{3}$$

is adopted here, with $m_e$ the mass of electron. It is worth to emphasize that $V_{\text{eff}}$ is introduced as a phenomenological short-range effective interaction and is not intended as a literal modification of the Coulomb interaction in vacuum. By construction, $V_{\text{eff}}(r)$ smoothly approaches the standard Coulomb form for $r \geqslant 10^3$ fm, so ordinary hydrogenic physics at atomic length scales is unaffected. In the present virial-consistency diagnostic, the role of $V_{\text{eff}}$ is to parameterize the additional short-range attraction required to balance the rapidly increasing relativistic kinetic-energy scale at femtometer radii. The implied enhancement relative to the point-Coulomb term at $r$ approximately a few fm within the adopted ansatz should therefore be interpreted as an effective strength (i.e., a constraint on any microscopic mechanism) rather than as a first-principles prediction of a new fundamental force.

To emphasize is that $V_{\text{eff}}$ (3) is not an interaction potential in the quantum–mechanical sense and is not used as a Hamiltonian operator defining eigenstates. Instead, the definition (3) is used only as an energy-dependent diagnostic ansatz, parametrizing a possible relativistic energy-balance for a charged particle in a strong static scalar potential, and used here only to test whether the virial-inspired matching procedure admits a stationary radius.

Reference [14] obtains a Schrödinger-like second-order equation by algebraic reduction of the Dirac equation in a static scalar Coulomb potential $\varphi(r)$. This reduces to $V_C(r) = -e\varphi(r)$ when $E \approx$ $m_e c^2$, $|e\varphi(r)| \ll m_e c^2$. In the relativistic regime, this transforms to Equation (3). It is woth to stress again that $V_{eff}$ serves as a phenomenological consistency test of whether ultra-relativistic kinematics combined with a central-field energy-balance ansatz may yield femtometer-scale solutions within the procedure used below. Figure 1 compares $V_C(r)$ and $V_{eff}(r)$.

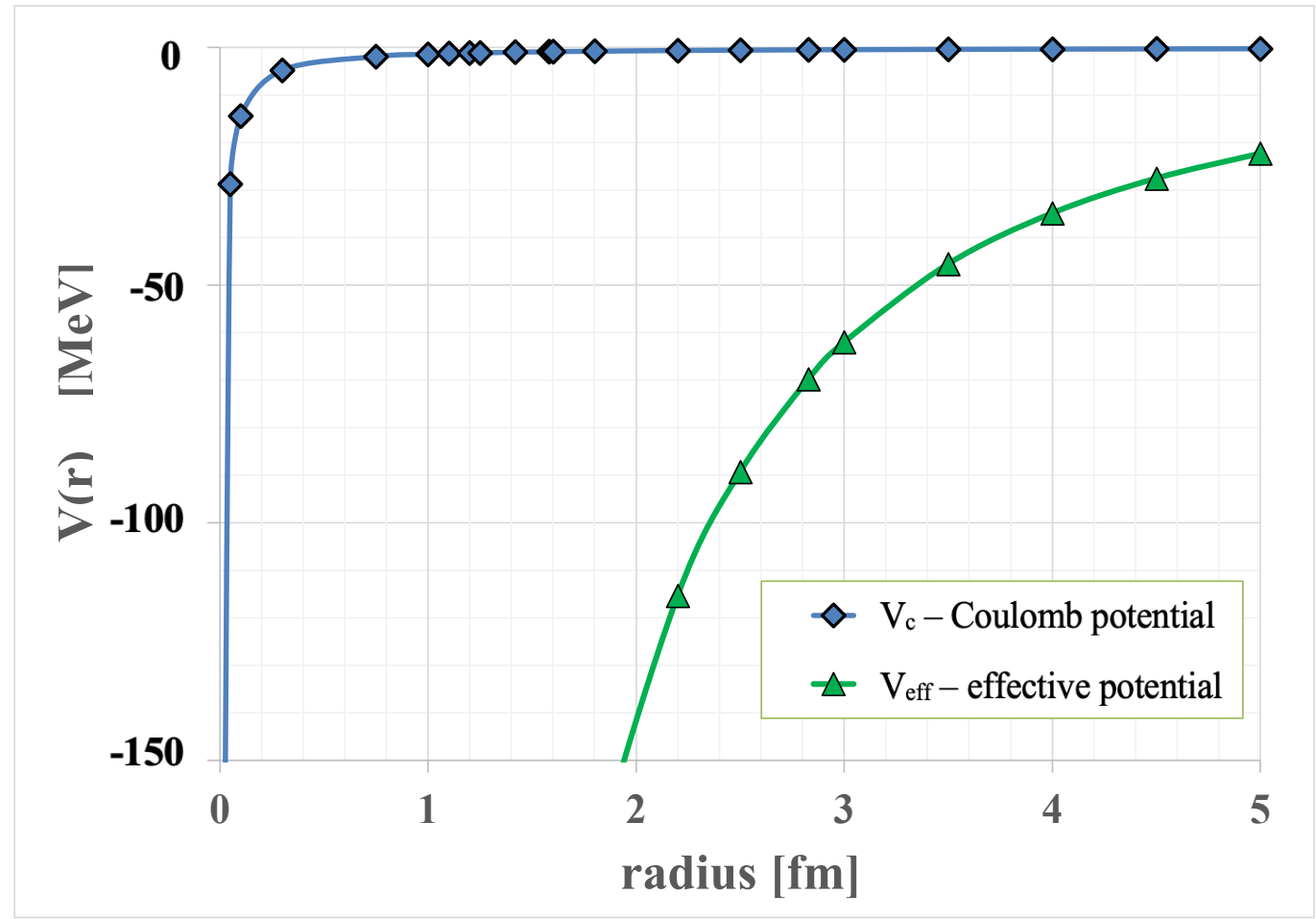

**Figure 1.** $V_C(r)$ and $V_{eff}(r)$ (3) potential versus. radius. Lines are to guide the eye.

### *3.2. Spin–Magnetic Interaction V(Spin.B)*

At radii of a few femtometers, the proton magnetic field is approximated here by its leading dipole term. This is used only to estimate the characteristic spin-dependent energy scale associated with proton magnetization. The spin–magnetic contribution is parametrized as

$$V_{(\mathrm{Spin}.B)}\,(r) = -\,\mu_e B_p(r) = (\mathrm{g_e}\mu_B/2\,\gamma_e)(\boldsymbol{\sigma}\cdot\boldsymbol{B}_p), \quad (4)$$

where $\boldsymbol{\mu}_e = -(\mathrm{g_e}\mu_B)\,\boldsymbol{\sigma}/2$ is electron dipole operator, $\boldsymbol{B}_p$ is magnetic field of proton, $g_e = 2.00232$ is the **electron g-factor**, and $\mu_B = e\hbar/2m_e$ is the Bohr magneton. The vector $\boldsymbol{\sigma}$ denotes Pauli spin matrices acting on the electron spin, related to the spin operator by $\mathbf{s} = (\hbar/2)\,\boldsymbol{\sigma}$. The factor $1/\gamma_e$ is introduced as a phenomenological Lorentz-scaling ansatz appropriate for an ultra-relativistic electron in the proton rest frame. The spin-magnetic congtribution changes sign depending on the relative orientation of the electron spin and the proton magnetic field, leading to a splitting between aligned and anti-aligned configurations.

The proton magnetic field is modeled as a dipole:

$$B_p(r) \approx (\mu_0/4\pi)\,2\mu_p/r^3, \quad (5)$$

where $\mu_0$ is the permeability of free space, $\mu_p = 2.793\mu_N$ is the magnetic moment of proton and $\mu_N = e\hbar/2m_p$ is the nuclear magneton, with $m_p$ the proton mass. At $r \approx 2.84$ fm, this gives $B_p \approx 1.2 \times 10^{11}$ T.

### *3.3. Spin–Orbit Interaction $V_{SO}$ (Order of Magnitude)*

For a central vector potential $V_C(r)$, the Foldy–Wouthuysen expansion of the Dirac equation yields the familiar spin–orbit coupling in the weak-filed regime [16]:

$$V_{\mathrm{SO}}(r) = \frac{1}{2m^2c^2}\,\frac{1}{r}\,\frac{dV_{\mathrm{C}}}{dr}, \quad (6)$$

where the Thomas factor is already included in the Dirac-based reduction.

At femtometer radii and ultra-relativistic electron energies, a consistent evaluation of spin–orbit effects requires a QED treatment including finite-size proton structure. In the present phenomenological framework, therefore Equation (6) is used just as a dimensional guide and a simple suppression estimate is applied by replacing $m_e \longrightarrow \gamma_e m_e$, which leads to $V_{SO} \approx 1/\gamma_e^2$. For $\gamma_e \approx 100$–$150$, this scaling implies that $V_{SO}$ is parametrically small compared to both the diagnostic term $V_{eff}$ and the spin–magnetic scale estimated in Section 2.2. As illustrated in Figure 2, the hierarchy of magnitude is $|V_{eff}| >> |V_{(Spin.B)}| >> |V_{SO}|$.

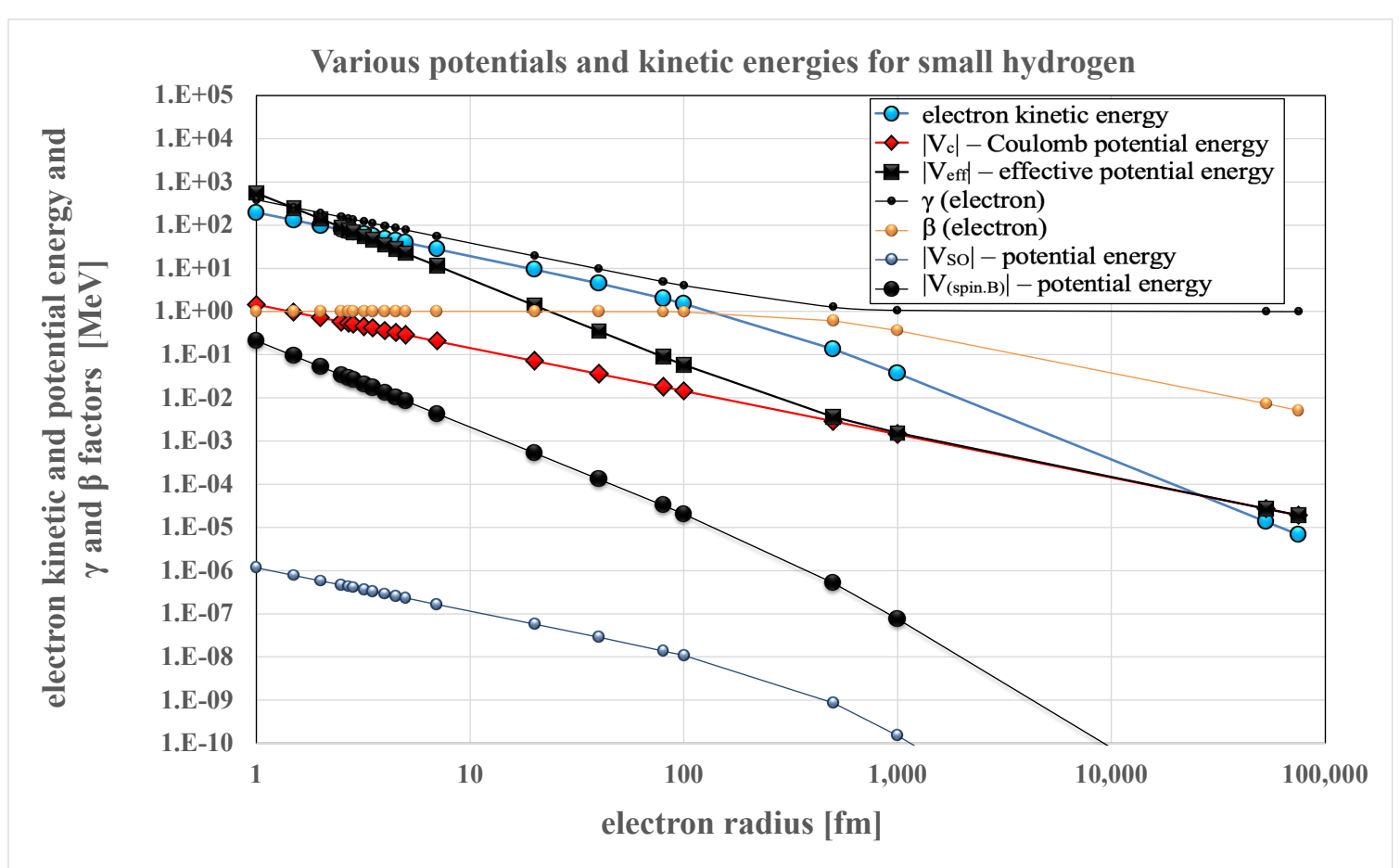

**Figure 2.** Comparison of the relativistic kinetic-energy scale $T_{kinetic}(r)$, with the magnitudes of the radius-dependent energy terms used in the diagnostic ansatz: $|V_C|$, $|V_{(Spin.B)}|$, $|V_{SO}|$, and $|V_{eff}|$. Here, "potential energy" denotes these radius-dependent energy contributions within the phenomenological framework. The plot shows the spin-aligned (positive $V_{(Spin.B\text{-}dipole)}$) configurationshows that the opposite spin orientation yields a corresponding slight shift of the compact solution toward smaller radii. Lines are to guide the eye. See text for more details.

It is worth to stress again that these terms are not components of a Hamiltonian, and no spectroscopic interpretation (for example, "fine structure") is implied. These features are quoted just to indicate the relative size of phenomenological spin-dependent energy contributions at femtometer radii within the limitations of the present framework.

## 4. Virial-Inspired Consistency Condition

Here, a virial-inspired local scaling test is applied: two radius-dependent energy scales, $T_{kinetic}(r)$ and $T_{virial}(r)$, are applied and a candidate consistency radius $r_c$ is defined by the matching condition $T_{kinetic}(r_c) = T_{virial}(r_c)$. This matching procedure is used only as a diagnostic for whether the adopted energy-balance ansatz admits self-consistent femtometer-scale radii under ultra-relativistic kinematics.

### 4.1. Electron Kinetic-Energy Scale $T_{kinetic}(r)$

Following de Broglie's standing-wave picture, a relation between momentum and radius is introduced by writing $2\pi r = n\lambda$, $p(r) = h/\lambda = \hbar\, n/r$, where $n$=1,2,... labels the number of de Broglie wavelengths fitting around a circular orbit, with $h$ the Planck constant. This relation is used here only as a phenomenological momentum–radius parametrization that sets the kinematic scale at a given radius.

The corresponding relativistic kinetic-energy scale is

$$T_{\text{kinetic}} = \sqrt{(p(r)\,c)^2 + (m_e c^2)^2} - m_e c^2. \quad (7)$$

For the ultra-relativistic regime $\gamma_e >> 1$, one may equivalently write $T_{\text{kinetic}}(r) = (\gamma_e - 1)\, m_e c^2$.

### 4.2. Virial-Inspired Energy Scale $T_{\text{virial}}(r)$

Writing $U(r) = \Sigma\, U_i(r)$ with local scaling $U_i(r) \propto r^{k_i}$,

$$T_{\text{virial}}(r) = \sum k_i\, [\,\gamma_e/(\gamma_e + 1)]\, U_i\,(\mathrm{r}) \quad (8)$$

is used [16–19]. Examples are as follows.

- For a Coulomb term $U_1 = V_C = -KZe^2/r$, the exponent is $k = -1$. Within the phenomenological relativistic virial parametrization of Equation (11), this gives $T_{\text{virial}} \longrightarrow -(½)V_C$ as $\gamma_e \longrightarrow 1$, and $T_{\text{virial}} \longrightarrow -V_C$ for $\gamma_e \longrightarrow \infty$.
- For a term $U(r) = 1/r^2$, $k = -2$, giving $T_{\text{virial}} \longrightarrow -2U$ as $\gamma_e \longrightarrow \infty$.
- For present model, $U(r) = V_{\text{eff}}(r) + V_{(\text{Spin}.B)}(r) + V_{SO}(r)$,

$$T_{\text{virial}} = \frac{\gamma^2}{\gamma+1}|V_C| + 2\frac{\gamma}{\gamma+1}\frac{{V_C}^2}{(2m_e c^2)} + + 3\frac{\gamma}{\gamma+1}\left|V_{(\text{spin}.B)}\right| - 3\frac{\gamma}{\gamma+1}V_{SO}. \quad (9)$$

The factor 3 reflects the local $1/r^3$ scaling of the magnetic-dipole and spin–orbit terms.

### 4.3. Local Matching Condition (Method A)

Let us define a candidate consistency radius $r_c$ by the matching condition

$$T_{\text{kinetic}}(r_c) = T_{\text{virial}}(r_c). \quad (10)$$

As an additional diagnostic, the total energy-balance functional $E_{\text{tot}}(r) = T_{\text{kinetic}}(r) + U(r)$ is defined and $E_{\text{tot}}(r_c) < 0$ is required, which is the ordinary energetic condition for a bound configuration.

Figure 3 illustrates this matching condition for ordinary hydrogen within a Bohr-model parametrization. $T_{\text{kinetic}}$ is evaluated from Equation (10) and $T_{\text{virial}}$ from Equation (11).

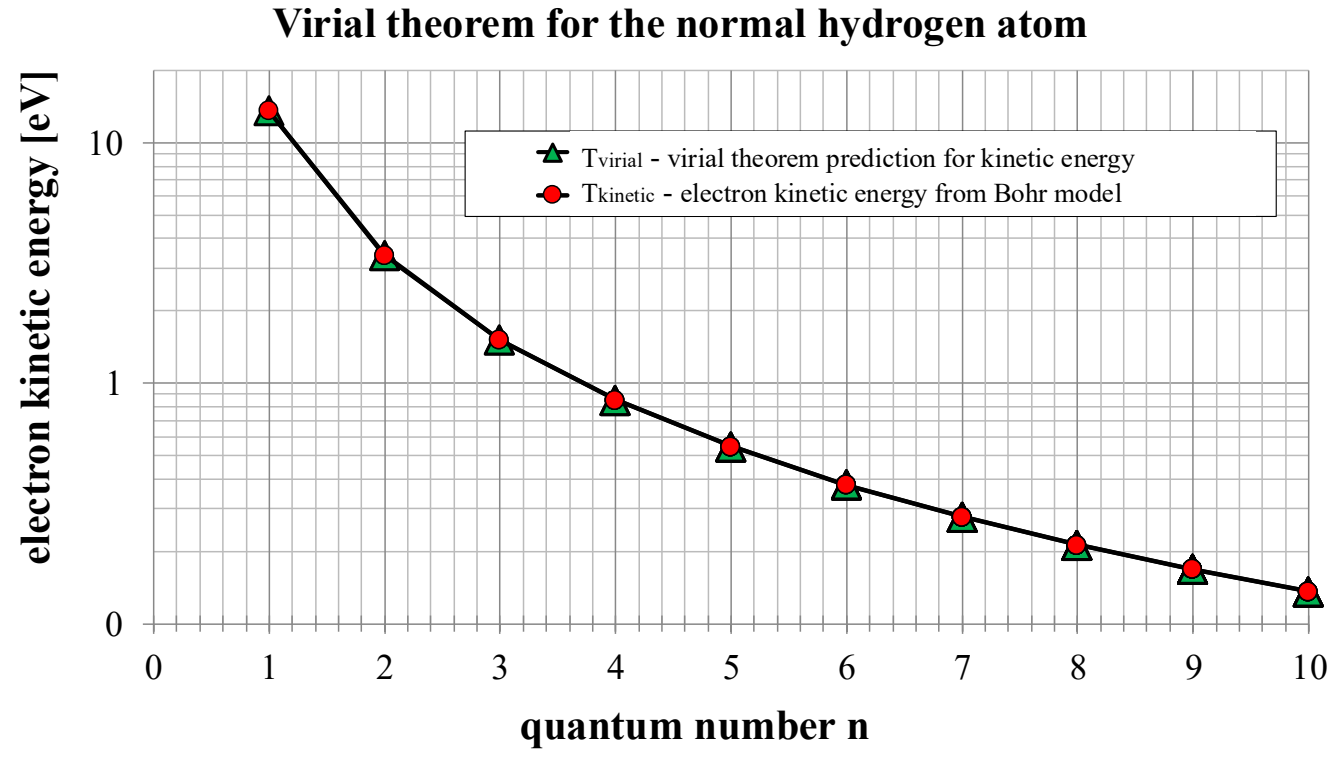

**Figure 3.** Local matching condition for ordinary hydrogen within the Bohr parametrization. $T_{\text{kinetic}}$ is evaluated from Equation (7) and $T_{\text{virial}}$ from Equation (8). Curves are to guide the eye.

### 4.4. Local Stationary Proxy (Method B: Cross-Check)

As a cross-check, let us evaluate the derivative-based local stationarity proxy [17] (see Appendix A):

$$< p \frac{\partial T_{\text{kinetic}}(p)}{\partial p} - r \frac{\partial U}{\partial r} > = 0, \tag{11}$$

where p is the magnitude of the momentum, r is the radial coordinate and U(r) depends only on scalar r. The angular brackets <…> denote the average over one period.

For a circular relativistic parameterization, Equation (11) becomes

$$\frac{(pc)^2}{\sqrt{((\boldsymbol{pc})^2 + (m_e c^2)^2}} = r \frac{\partial U}{\partial r}. \tag{12}$$

Here $U(r) = V_{\text{eff}} + V_{(\text{Spin}.B)} + V_{\text{SO}}$. Solving Equation (12) yields radii that provide an internal consistency check of the matching solution obtained from Equation (A8).

## 5. Results

If a compact e–p configuration exists in nature, it must correspond to a dynamically stable (or metastable) bound state. The present analysis does not address formation or lifetime; it only identifies candidate stationary radii within a virial-inspired energy-balance test.

### 5.1. Coulomb Potential $V_C$

Applying the virial-consistency procedure to the entirely Coulomb potential $V_C = -KZe^2/r$ recovers the ordinary hydrogen solution at the Bohr radius, but no additional femtometer-scale solution is found. Finite-size Coulomb forms (Smith–Johnson, Nix forms [4,5]) do not change this conclusion.

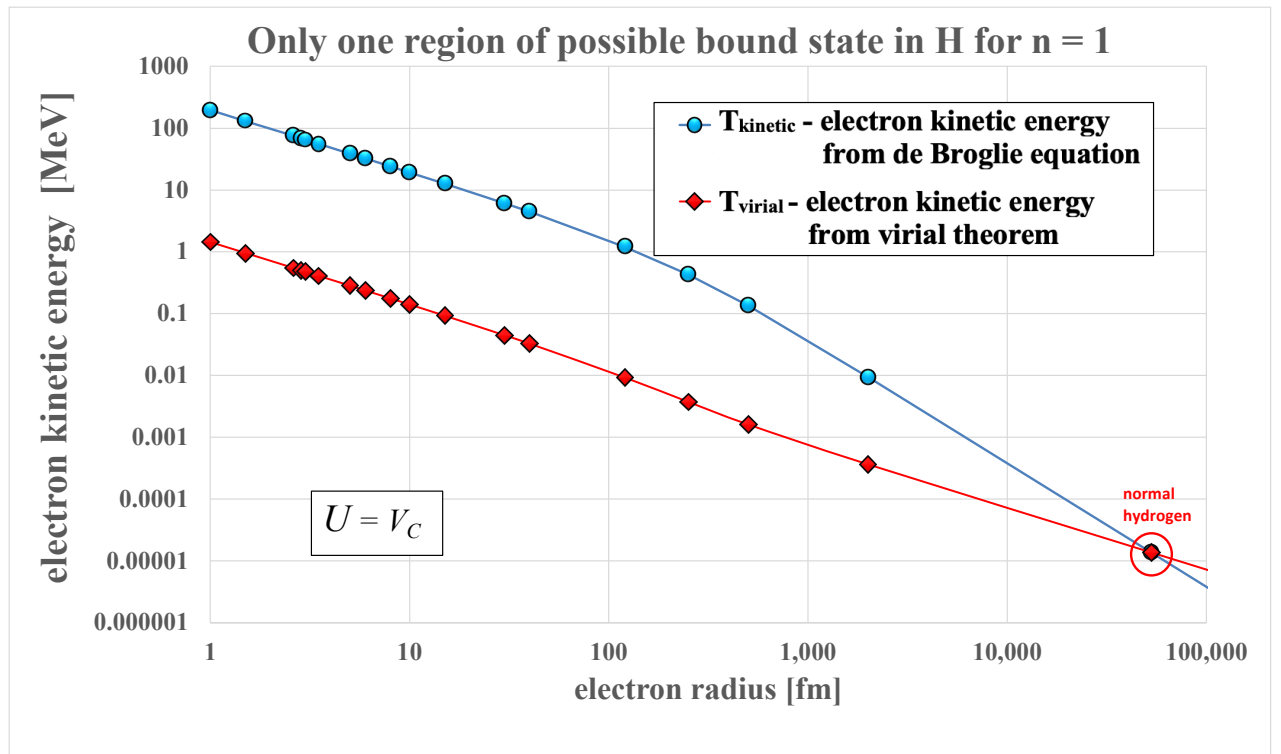

**Figure 4.** Virial-consistency test applied to the entirely Coulomb interaction in the e–p system. The method reproduces the known stable Bohr-radius solution of ordinary hydrogen, but no additional femtometer-scale matching radius appears. Curves are to guide the eye.

### 5.2. Effective Potential $V_{\text{eff}}$

Introducing the semi-relativistic effective potential $V_{\text{eff}}$ (3) yields a compact femtometer-scale matching radius near $r \approx 2.8$ fm. Figure 5 shows two radii at which the matching condition is satisfied: one at the Bohr scale (ordinary hydrogen) and a second at $r \approx 2.84$ fm. Table 1 lists the corresponding quantities. The total energy-balance functional $E_{\text{tot}} = T_{\text{kinetic}} + U$ becomes negative at this radius, indicating a negative-energy configuration within the adopted ansatz. The entries with $n = 2$ illustrate robustness with respect to the de Broglie parametrization; in the following, the $n$ =

1 case is considered. The quantity $E_{BE}$ in Table 1 is defined here only as an energy-balance deficit: $E_{BE} = T_{kinetic} - |U|$.

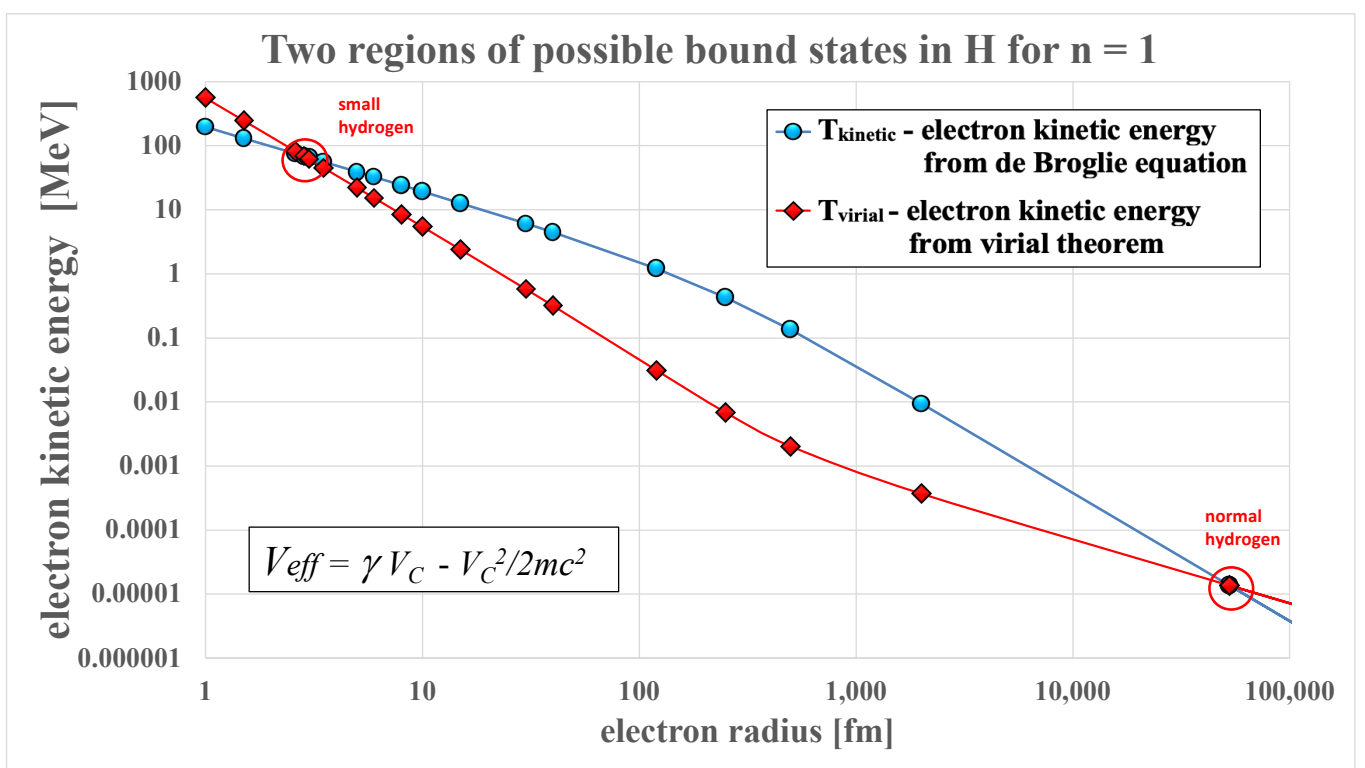

**Figure 5.** Virial-consistency solutions showing two stability regions—one for ordinary hydrogen and one for small hydrogen—calculated for $V_{eff} = \gamma V_C - V_C^2/2\,mc^2$. Curves are to guide the eye.

**Table 1.** Compact stationary virial-consistent solutions obtained using $V_{eff.}$ (3).

| $r_c$ [fm] | $U = \gamma V_C - V_C^2/2\,mc^2$ [MeV] | $T_{kinetic}$ [MeV] | $M(pe^-)$ * [MeV/$c^2$] | $E_{BE}$ ** [keV] |
|---|---|---|---|---|
| 2.8386 | −69.275 | 69.016 | 938.524 | −259.3 |
| 2.8283 | −139.371 | 139.115 | 938.527 | −255.7 |

* Mass at $r_c$: $M(pe^-) = m_p + \gamma m_e - |U|$. ** Energy-balance deficit: $E_{BE} = T_{kinetic} - |U|$. See text for details.

### *5.3. Combined Potential $U = V_{eff} + V_{(Spin.B)} + V_{SO}$*

The virial-stability analysis is applied using the full effective potential $U = V_{eff} + V_{(Spin.B)} + V_{SO}$. Unless otherwise stated, all figures in this section correspond to the spin-aligned configurateion ($+V_{(Spin.B)}$). Figure 6 compares $T_{kinetic}(r)$ and $T_{virial}(r)$ evaluated from Equations (10)–(12). Two radii satisfy the matching condition $T_{kinetic}(r) = T_{virial}(r)$: one corresponding to ordinary hydrogen and a second corresponding to a compact femtometer-scale radius. Thus, adding the spin–magnetic term does not significantly shift the compact matching radius.

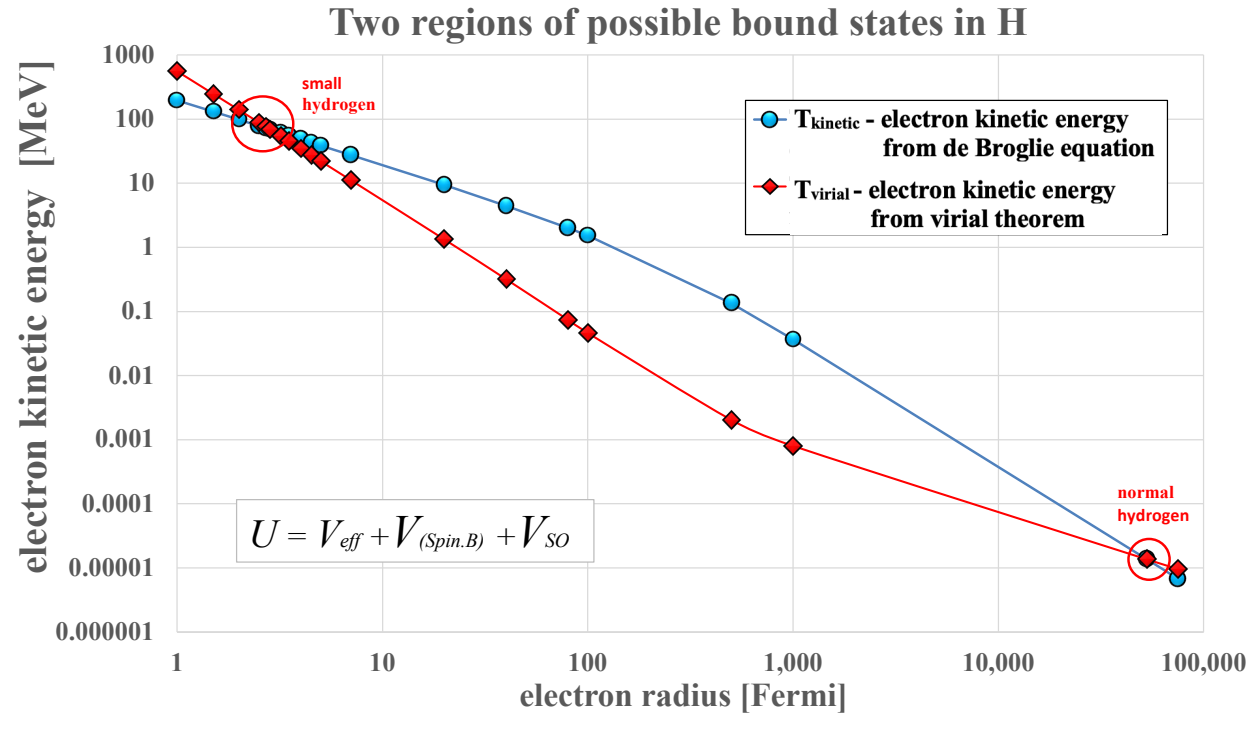

**Figure 6.** Comparison of $T_{\text{kinetic}}$ and $T_{\text{virial}}$ using the full potential $V_{\text{eff}} + V_{(\text{Spin}.B)} + V_{\text{SO}}$. Two regions of stability appear: one corresponding to ordinary hydrogen and one at small radius. Curves are to guide the eye.

To check internal consistency, the derivative-based stationarity proxy (Method B, Equation (12)) is applied. Figure 7 exhibits a sharp minimum at $r \approx 2.84$ fm. The dip is narrow (full width at half maximum, FWHM, of about 0.5 fm), reflecting a sharply defined extremum of the chosen diagnostic function within this procedure. This width should not be interpreted as a physical localization length or a dynamical stability measure. Table 2 lists characteristic quantities associated with the compact e-p solution obtained within the phenomenological model.

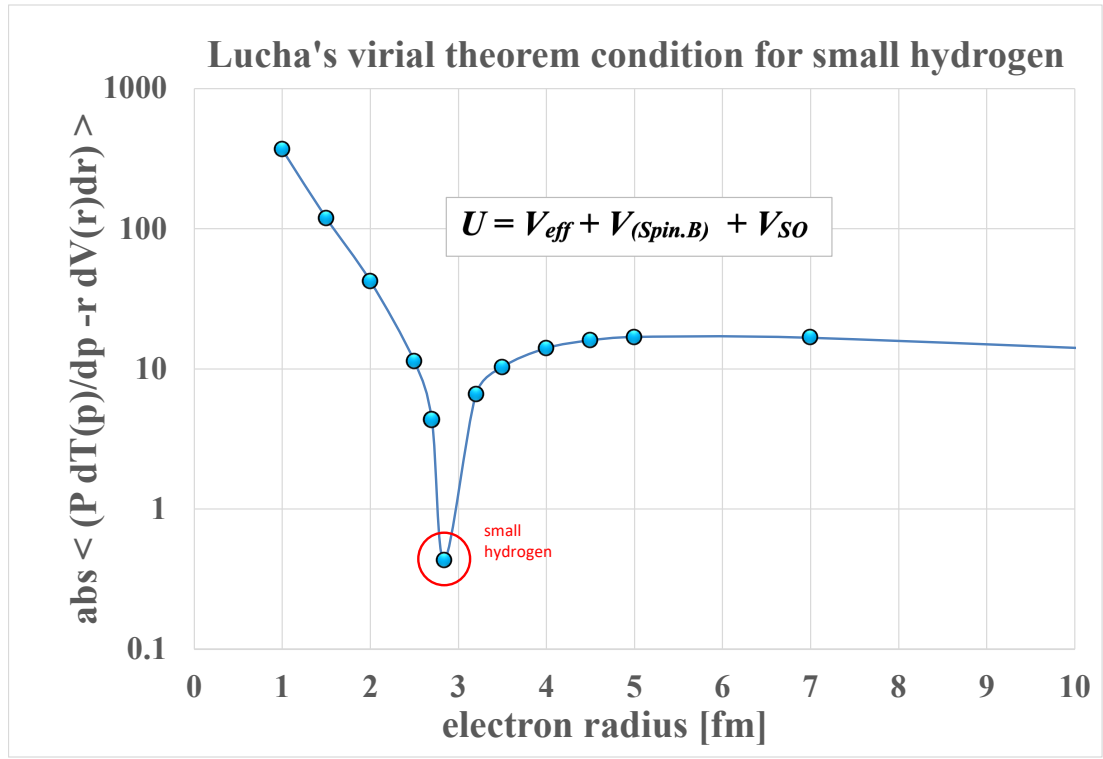

**Figure 7.** Solution of the relativistic virial equation (Method B). A sharp dip at $r \approx 2.84$ fm confirms astable compact solution. Curves are to guide the eye. See text for details.

**Table 2.** Small hydrogen solution using potential $U = V_{\text{eff}} \pm V_{(\text{Spin}.B)} + V_{\text{SO}}$.

| Quantity | Spin (+) | Spin (-) |
| --- | --- | --- |
| $r_c$ [fm] | 2.8419 | 2.8355 |
| $U(r)$ [MeV] | −69.137 | −69.399 |
| $T_{\text{kinetic}}$ [MeV] | 68.933 | 69.091 |
| $T_{\text{virial}}$ [MeV] | 68.933 | 69.091 |
| $B_p$ [T] | $1.229 \times 10^{11}$ | $1.237 \times 10^{11}$ |
| $V_{(\text{Spin}.B)}$ [MeV] | −0.0262 | +0.0263 |
| $V_{\text{eff}}$ [MeV] | −69.111 | −69.425 |
| $V_{\text{SO}}$ [MeV] | $+4.06 \times 10^{-7}$ for $\ell$=1, $s$=1/2<br>$-8.15 \times 10^{-7}$ for $\ell$=1, $s$=−1/2<br>0 for $\ell$=0 | same as spin + |
| $E_{\text{BE}}$ [keV] | −204 | −308 |
| mass [MeV/$c^2$] | 938.58 | 938.48 |

$E_{\text{central}}$ is defined as the energy scale obtained using the effective potential $V_{\text{eff}}$ alone, which gives $E_{\text{BE}} \approx$ -260 keV (Table 1). When the full potential $U = V_{\text{eff}} \pm V_{(\text{Spin}.B)} + V_{\text{SO}}$ is included, two stationary solutions are obtained, with energy-balance deficits of approximately –204 keV and –308 keV (Table 2). This corresponds to a characteristic spin-dependent splitting of approximately 100 keV. The spin–magnetic term introduces a local contribution of magnitude $|V_{(\text{Spin}.B)}| \approx 25$ keV at the compact consistency radius (Table 2). Within the present virial-consistency framework, however, this term shifts the stationary solution in opposite directions for the two spin orientations, thereby producing this approximately 100 keV splitting. These values should be interpreted as indicative energy scales within the adopted phenomenological framework.

The interpretation is as follows.

- Within the adopted energy-balance ansatz, $V_{eff}$ sets the dominant central energy scale (about 260 keV).
- The spin–magnetic contribution leads to two stationary solutions, corresponding to a characteristic splitting of order $\Delta E_{spin} \approx 100$ keV.
- The spin–orbit term $V_{SO}$ is suppressed by $(1/\gamma_e)^2$ and contributes only $|V_{SO}| \approx 10^{-7}$ MeV, negligible for binding.
- The associated mass scale inferred within the model lies below $m_p + m_e$ and below the neutron mass. No claim is made regarding dynamical stability, decay channels, or formation probability.
- Earlier models (i.e., early DDL-based work) suggested binding energies of order 0.5 MeV and a possible connection to the Galactic 511 keV line. In the present analysis, the characteristic energy scales lie in the range of 200-310 keV, centered near 260 keV, and therefore are not consistent with such an interpretation.

For comparison, the hyperfine splitting of ordinary hydrogen is $5.879 \times 10^{-6}$ eV (the 21 cm line), reflecting the much weaker magnetic field and nonrelativistic kinematics at atomic length scales.

Table 3 summarizes typical parameters of small and ordinary hydrogen. As Table 3 shows, mass $M(pe^-)$ lies below $m_p + m_e$ and below $m_n$.

**Table 3.** Small hydrogen properties.

| Quantity | Ordinary Hydrogen | Small Hydrogen |
|---|---|---|
| $n$ | 1 | 1 |
| electron radius | 0.529 A | 2.84 fm |
| electron de Broglie wavelength | 3.322 A | 17.86 fm |
| electron de Broglie wave frequency | $\approx 6.6 \times 10^{15}$ Hz | $\approx 1.68 \times 10^{22}$ Hz |
| electron $\beta = v/c$ | $\approx 7.3 \times 10^{-3}$ | $\approx 0.9999729$ |
| electron $\gamma_e$ | 1.0000266 | $\approx 136$ |
| total mass $M(pe^-)$ | 938.78 MeV/$c^2$ | 938.58 MeV/$c^2$ |

One may speculate that analogous consistency arguments could be explored for other nuclei, although no such analysis is attempted here and no conclusions regarding multi-nucleon systems are drawn.

The author does not claim to establish the existence of small hydrogen as a physical particle, nor to replace a full two-body QED treatment. It is just shown that within a virial-consistency diagnostic incorporating relativistic kinematics and finite-size effects, a compact femtometer-scale stationary e–p configuration is not obviously inconsistent energetically. Its existence is therefore an empirical question.

## 6. Heisenberg Uncertainty Principle

The spatial confinement of an electron to femtometer scales ($\Delta x \lesssim$ 2–3 fm) implies a momentum uncertainty $\Delta p \gtrsim \hbar/(2\Delta x) \approx$ 100–200 MeV/c. Such momenta correspond to electron Lorentz factors $\gamma_e \approx$ 100–150, placing the system firmly in the ultra-relativistic regime. This is consistent with the relativistic assumptions employed throughout the virial analysis.

Within the phenomenological framework adopted here, these large kinetic-energy scales are compared against the characteristic energy scales associated with the diagnostic central-field terms introduced in Section 2. The uncertainty principle therefore does not, by itself, rule out compact stationary electron–proton configurations; rather, it emphasizes that any such configuration—if dynamically realizable—must lie far beyond the nonrelativistic Coulomb regime and requires ultra-relativistic dynamics.

A definitive assessment of whether such relativistic confinement can occur in a real two-body system requires a full quantum-field-theoretic treatment incorporating QED effects and proton structure. The present study does not attempt such a derivation, but instead addresses the more limited question of whether the uncertainty principle alone constitutes a fundamental obstruction to femtometer-scale electron–proton configurations within this diagnostic energy-balance approach.

## 7. Interactions of Small Hydrogen

Because small hydrogen is electrically neutral and extremely compact ($r \approx$ 3 fm), its electromagnetic scattering cross-section in matter is expected to be exceptionally small. At astrophysically relevant velocities, its kinetic energy per atom is typically only tens to hundreds of keV (for example, about 105 keV at 4500 km/s), insufficient to ionize the bound electron or cause nuclear disruption. Thus, if such objects exist, they would be expected to deposit very little energy per unit length in ordinary matter and to interact weakly through electromagnetic processes, with gravity dominating their large-scale dynamics.

At thermal energies, however, the absence of a large Coulomb barrier may allow small hydrogen to be captured by positively charged nuclei. The cross-section for such capture is not presently known and believed to depend on short-range QED and nuclear structure. At collision energies comparable to or exceeding the binding energy, the atom may be ionized; at GeV-scale energies, it would behave similarly to a neutron in initiating hadronic cascades.

These considerations are qualitative and do not constitute a prediction of interaction rates or abundances.

## 8. Can Small Hydrogen Atoms Be Detected?

Formation of a compact *e*–*p* configuration at femtometer scales would require (i) a matched relative velocity between the proton and electron and (ii) an electron de Broglie wavelength comparable to a few femtometers. Such conditions are expected only in environments involving ultra-relativistic charged particles (for example, the early Universe, relativistic plasmas, strong shocks, or compact-object magnetospheres), although a quantitative formation rate is not known.

If such objects to exist today, laboratory searches may to rely on their interactions with nuclei. Because the configuration is electrically neutral and highly compact, it actually may penetrate atomic electron clouds and approach nuclei without experiencing the large long-range Coulomb barrier characteristic of charged projectiles. Whether this leads to appreciable capture or nuclear effects is uncertain and would depend on short-range QED dynamics and nuclear structure.

For certain targets (for instance, boron), one may speculate that capture of a compact e–p configuration could alter nuclear stability and produce identifiable secondary products (for example, $\alpha$-emission). At present, however, no established experimental signature or confirmed detection exists, and the relevant cross sections and branching ratios remain unknown.

Additionally, if a nucleus were to bind such a compact e–p configuration in a persistent way, the outer electronic structure could resemble that of an atom with effective charge ($Z$−1), suggesting a possible spectroscopic search strategy.

## 9. Astrophysics Implications

If compact neutral e–p configurations exist in nature, the phenomenological energy scales inferred in Sections 2–4 suggest that electromagnetic emission or absorption features—if any radiative channels exist—would plausibly fall in the 100–400 keV range. In particular, the characteristic spin-dependent scale is of order $\Delta E_{\text{spin}} \approx 100$ keV, while the central energy-balance deficit associated with the compact matching solution is of about 260 keV.

No confirmed astrophysical feature has been identified with such a hypothetical signal, and the present work does not predict an actual transition rate, branching ratio, or line strength. The main observational implication is therefore limited to a suggested search window: sensitive measurements in the about 100–400 keV range, with good control of instrumental backgrounds, could constrain or exclude narrow excesses associated with any compact neutral e–p component, if to exist and radiate.

### *9.1. Galactic Rotation Curves and Cosmic Time Evolution*

It is worth to emphasize that the following discussion is qualitative and is not intended as an explanation of galaxy rotation curves, nor as an alternative to nonbaryonic dark matter. The purpose is only to note that if a weakly interacting baryonic component were produced predominantly in energetic stellar environments, its abundance could increase with cosmic time.

In this context, it is of interest that Ref. [20] reports that massive star-forming disk galaxies at redshifts $z \approx 0.6$–$2.6$ appear largely baryon-dominated within their observed radii, with rotation curves that decline at large radius, whereas many present-day ($z \approx 0$) galaxies exhibit extended nearly flat rotation curves commonly interpreted in terms of dark-matter halos. This phenomenology does not uniquely favor any particular explanation, but it illustrates that the "dark component" inferred in galaxies is not trivially constant in appearance across cosmic time and environments.

Likewise, systems such as NGC 1277 have been discussed as cases with comparatively little evidence for dark matter within the optical radius [21]. Whether such diversity reflects baryonic feedback, halo contraction/expansion, selection effects, or other physics remains an active topic. In the present paper, this point is mentioned only to highlight that galaxy-scale phenomenology is complex, and therefore, any proposed compact baryonic component—if it existed—would require attentive modeling rather than simplistic claims.

### *9.2. Early-Universe Constraints (BBN)*

If compact e-p configurations were produced in the early Universe with non-negligible abundance, they would be constrained by Big Bang nucleosynthesis (BBN), since BBN successfully accounts for the observed light-element abundances at approximately the percent level. Any scenario in which such objects form before or during BBN must therefore be strongly limited in abundance (at most at about the 1% level as an approximate benchmark [22], depending on how the objects participate in nuclear reaction networks).

A quantitative treatment of formation channels and BBN reaction impacts is beyond the scope of this phenomenological study; BBN provides a strong consistency constraint that any early-Universe production scenario would need to satisfy.

### *9.3. Cluster Collisions (Bullet Cluster)*

The Bullet Cluster provides a known testbed for dark-matter phenomenology, in particular the apparent separation between the X-ray-emitting intracluster gas and the gravitational lensing mass [23,24].

If compact neutral e–p configurations exist, their rest mass would be approximately $M(pe^-) \approx 938.6$ MeV/$c^2$, and for a typical cluster collision speed of about 1310 km/s, the kinetic energy of such an object is $T = \frac{1}{2} mv^2 \approx 10$ keV. This energy is an order of magnitude lower than the 200-300 keV binding energy inferred from the virtual analysis discussed in Sections 2–4, so such collisions are unlikely to ionize the compact e-p system or produce any radiative signature. In that case, the component passes through the hot gas with no necessary radiative signature, while remaining detectable through gravity (lensing)—qualitatively similar to the collisionless behavior inferred in cluster mergers.

### *9.4. Possible Connection to INTEGRAL MeV Lines*

Many $\gamma$-ray lines seen by INTEGRAL/SPI instrument in approximately the MeV range are commonly attributed to activation and neutron-capture processes in spacecraft and detector materials. The INTEGRAL Collaboration has emphasized that reproducing the full richness and relative strengths of instrumental background lines is challenging in Monte-Carlo simulations, reflecting the complexity of activation processes and surrounding material [25].

This motivates the general value of future instruments optimized for low activation (minimal surrounding mass, low-activation materials, and an orbit far from strong activation environments) when searching for weak narrow features. In the present context, the point is not that INTEGRAL instrument observed any small-hydrogen feature, but that high-resolution directional search, low-background instrumentation would be the appropriate way to constrain narrow spectral features in about the 100–400 keV band if one wishes to test the energy scales suggested by the present paper.

The astrophysical remarks above are included only as qualitative context and as potential sources of constraints; they do not constitute evidence for compact e-p configurations, and do not replace the need for laboratory tests and a full two-body QED/QFT (quantum field theory) treatment.

## 10. Conclusions

In this study, a relativistic, virial-inspired energy-balance analysis was used to test whether an electron confined to femtometer radii can remain self-consistent at the level of relativistic energy balance in an e–p system. Within the adopted phenomenological ansatz—including finite-size regularization and approximate spin-dependent contributions—the procedure yields a compact stationary radius at the few-femtometer scale. While this does not establish a physical bound state, it shows that relativistic kinematics and simple stationarity constraints do not, by themselves, exclude such a compact configuration within the assumptions of the model.

Within this framework, the dominant central energy scale associated with the compact solution is of about 260 keV, while spin-dependent contributions introduce a characteristic scale of about 100 keV. As soon as these values are not obtained from a two-body Hamiltonian eigenvalue problem or a first-principles QED bound-state calculation, they should be interpreted as indicative energy scales emerging from the consistency analysis, rather than as predictions of spectroscopic levels or transition energies.

The present study is intended as a consistency diagnostic rather than a microscopic theory. Determining whether such configurations exist in nature—and, if so, their formation probability, lifetime, decay channels, and dynamical stability—ultimately requires a full two-body treatment incorporating QED dynamics and proton structure.

If compact neutral e–p configurations exist and possess radiative or interaction channels, the keV-scale energies identified here motivate targeted laboratory and astrophysical searches in approximately the 200–310 keV range. Appendix B outlines illustrative experimental strategies that could constrain or test this possibility.

## Appendix A: Relation to Lagrangian, Hamiltonian, and Virial Formulations

This Appendix provides brief motivation for the stationarity criteria used in the paper. The relations quoted are standard for a relativistic particle in an effective central potential and are used here only as heuristic consistency guides.

### *A1. Relativistic Lagrangian in a Central Potential*

For a particle of mass m moving in a spherically symmetric effective potential *U(r)*, the relativistic Lagrangian may be written as

$$L(r, v) = -\frac{mc^2}{\gamma_e} - U(r), \tag{A1}$$

with the Lorentz factor $\gamma_e = 1/\sqrt{1-(v/c)^2}$, position $r$, and velocity $v$ of a partice of mass $m$.

The action is

$$S = \int L(r, v) dt \tag{A2}$$

and the physical trajectories satisfy the stationary-action condition $\delta S$ = 0, which yields the relativistic equations of motion in *U(r)*.

### *A2. Virial Relation for Bounded Orbits*

For bounded/periodic motion in a central potential, the relativistic virial relation follows as a standard consequence of the equations of motion when averaged over one period (under the standard assumptions of periodicity and finite time averages):

$$< \boldsymbol{p} \cdot \boldsymbol{v} > = < r \frac{dU}{dr} >. \tag{A3}$$

For circular motion (or for a circular-orbit parametrization used as a representative stationary configuration), this motivates a local stationarity condition relating the kinetic-energy scale to the radial scaling of *U(r)*. In the paper, this point is implemented through the matching of the radius-dependent diagnostic scales $T_{\text{kinetic}}(r)$ and $T_{\text{virial}}(r)$.

### *A3. Hamiltonian Formulation and Local Stationrity*

The canonical momentum is

$$\boldsymbol{p} = \frac{\partial L}{\partial \boldsymbol{v}} = m\gamma_e \boldsymbol{v}. \tag{A4}$$

where $\boldsymbol{v}$ is velocity vector.

The corresponding Hamiltonian is

$$H = \boldsymbol{p} \cdot \boldsymbol{v} - L = \gamma_e mc^2 + U(r), \tag{A5}$$

i.e., the total energy functional for motion in an effective central field. For circular motion (or equivalently for vanishing radial acceleration), the equilibrium radius satisfies the stationarity condition

$$\frac{\partial \mathrm{H}}{\partial \mathrm{r}}\Big|_{\mathrm{r}=\mathrm{r_c}} = 0, \tag{A6}$$

which yields

$$\frac{p^2c^2}{\sqrt{p^2c^2 + m^2c^4}} = r\,\frac{dU}{dr}. \tag{A7}$$

This is the derivative-based stationarity proxy used in the main text (Method B) as an internal cross-check.

## Appendix B: Illustrative Experimental Considerations for Compact Electron–Proton Configurations

This Appendix outlines qualitative considerations for how compact electron–proton configurations, if to exist, might be probed experimentally. No claim is made regarding formation probability, production cross section, lifetime, or experimental feasibility. The discussion is intended solely to illustrate possible energy scales and kinematic handles suggested by the virial analysis presented in the main text.

### *B1. Analogy with Ordinary Hydrogen Formation*

The formation of ordinary hydrogen provides a useful conceptual analogy. Ordinary hydrogen forms when the electron and proton move slowly relative to each other.

By analogy, formation of a compact electron–proton configuration would require matched velocities and a de Broglie wavelength comparable to the femtometer-scale radius inferred from the virial analysis. Whether such conditions occur in nature or in laboratory environments remains an open question.

The same considerations are invoked here solely as a qualitative analogy for the formation of a compact electron–proton configuration.

### *B2. Co-Moving Electron–Proton Configurations*

Tables 2 and 3 summarize the characteristic of the compact e–p state:

- radius $r_c \approx 2.84$ fm,
- de Broglie wavelength $\lambda \approx 17.9$ fm,
- electron kinetic energy $T_{\mathrm{Kinetic}} \approx 68.933$ MeV, and
- electron Lorentz factor $\gamma_e \approx 135.9$ and $\beta = 0.9999729$.

In this illustrative framework, formation would require the proton and electron to have approximately matched velocities. For $\beta = 0.9999729$, the corresponding proton total energy is $E_p \approx$ 127.51 GeV.

Thus, one kinematically consistent configuration would involve co-propagating relativistic electron and proton beams, as sketched in Figure A1a. In the rest frame of such a co-moving e–p pair, the model parameters imply that formation might be accompanied by emission of radiation with a characteristic rest-frame energy $E_{Source} \approx$ 100–150 keV.

Doppler-boosted gamma signal in the laboratory frame, the photon energy is

$$E_{\text{Observed}} = E_{\text{Source}}/[\,\gamma\,(1 - \beta \cos f)], \tag{A8}$$

where $f$ is the angle between the photon direction and the beam axis (Figure A1b).

- At $f \approx 6.95^{\circ}$, $E_{\text{Observed}} \approx E_{Source}$.
- At $f = 0^{\circ}$, $E_{\text{Observed}} \approx$ 40.77 MeV, a value highly sensitive to relatively small uncertainities in $\gamma$, and therefore unsuitable for a precision line search.

Therefore, a practically favorable angular region is approximately 5–20°, where the Doppler factor is modest and the observed energy remains close to the intrinsic about 100–150 keV scale. This kinematic relation is included only to illustrate how a rest-frame energy scale may transform to the laboratory frame under relativistic motion, and does not imply a predicted signal strength.

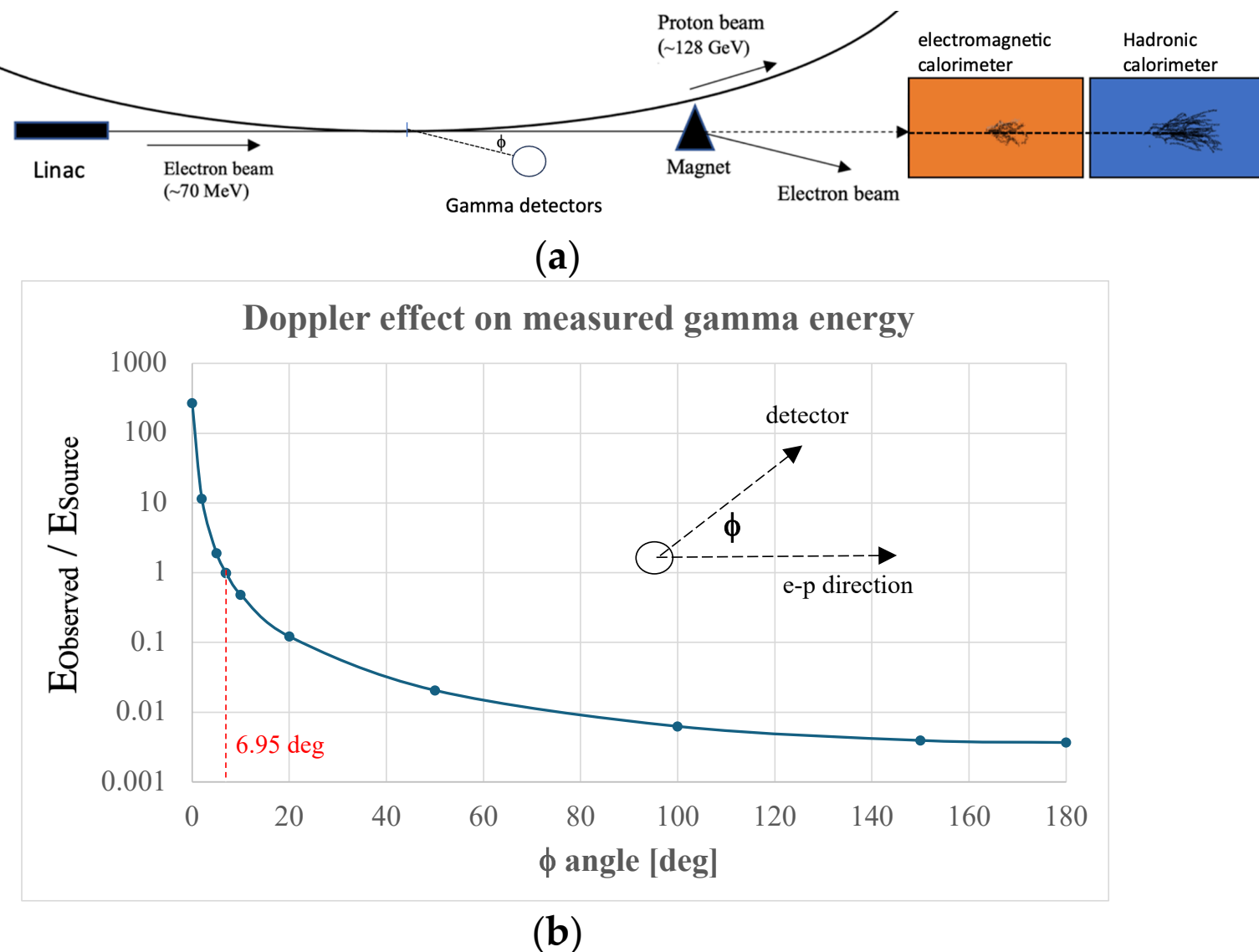

**Figure A1.** (**a**) Illustrative co-moving *e–p* geometry showing one kinematically consistent way to match relativistic velocities, motivated by the values in Tables 2 and 3. (**b**) Doppler transformation of a rest-frame energy scale under relativistic motion (Equation (14)). Vertical dash line at 6.95° corresponds to $E_{\text{Observed}} = E_{\text{Source}}$. Curves are to guide the eye.

Experimental Configuration

If a compact e–p configuration was produced and survived to reach a detector, its subsequent interaction in a calorimeter would plausibly result in

- one electromagnetic shower (electron),
- one hadronic shower (proton),

nearly collinear because of the large Lorentz boost. Their combined energy is be expected to be near the inferred mass, about 938.58 MeV/$c^2$.

Facilities with relativistic co-moving electron and proton beams, such as BNL (Berkeley Nationa Laboratory), Fermilab (Fermi National Accelerator Laboratory), or CERN (the European Organization for Nuclear Research), could in principle explore such kinematics.

### *B3. Search in High-Energy Collisions*

Large collider experiments (BaBar, Belle II, LHCb, CMS, ATLAS, or the EIC experiments) may actually place constraints on or search for experimentally accessible signatures consistent with a compact e–p configuration produced in high-energy interactions:

1. missing-mass structure near about 938.6 MeV/$c^2$ in exclusive or semi-inclusive final states;
2. channels consistent with a neutral, compact e–p system, if sufficiently long-lived;
3. absence of a displaced vertex, if the lifetime is considerably short;
4. a V-shaped decay topology (Figure A2a), if the lifetime is measurable;
5. two nearly collinear calorimeter showers in the calorimeter (Figure A2b), one electromagnetic and one hadronic.

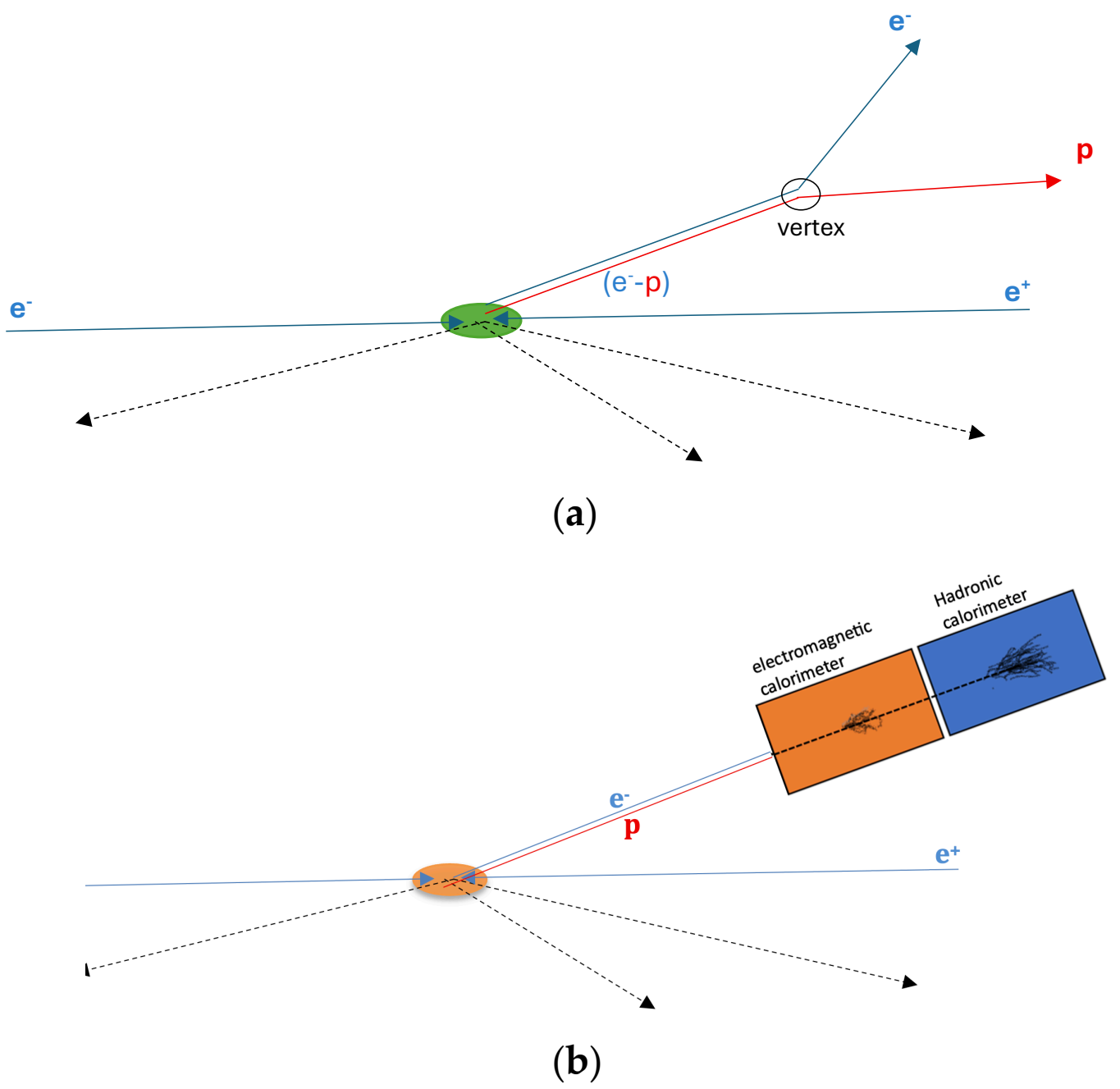

(**a**)

(**b**)

**Figure A2.** (**a**) Illustrative decay topology that could arise if a compact electron–proton configuration has a measurable lifetime. (**b**) Schematic calorimetric signature that could occur if such a configuration were stable on detector scales.

### *B4. Satellite-Based Search*

A dedicated low-activation space instrument can provide strong constraints on narrow spectral features in the relevant energy range. Key capabilities include:

- sensitivity to narrow features near the spin-dependent scale of approximately 100 keV,
- sensitivity to features near the central scale of about 260 keV, with possible spin-dependent shift toward about 200 and about 310 keV within the adopted ansatz,
- minimal nuclear-activation backgrounds,
- directional capability, and
- ability to distinguish astrophysical variability from instrumental effects.

Experience from INTEGRAL/SPI instrument shows that activation lines are quite complicated to model precisely. A purpose-built low-mass detector, placed far from the Earth–Sun system, would reduce activation backgrounds and enable a cleaner search for narrow features in the approximately 100–400 keV band.

## Appendix C: Speculative Astrophysical Conditions for Small-Hydrogen Formation

This Appendix does not propose or model the formation of compact electron–proton hypothetical bound states in astrophysical environments, nor does it interpret existing observations as evidence for their existence. Its purpose is solely to identify environments in which the characteristic frequency and energy scales inferred from the virial analysis would not be a priori forbidden, and to indicate observational regimes that therefore constrain such scenarios.

George Schott [26] showed that a spatially extended charge distribution undergoing periodic motion does not radiate (this is mentioned only as a speculative classical analogy for standing-wave configurations and is not used in the analysis) when its characteristic size is commensurate with the oscillation period, even for non-circular or three-dimensional motion; related results were verified in later studies [27–29]. Although such classical models do not represent a fundamental description of atomic structure, they provide an instructive heuristic analogy for bound systems governed by standing-wave conditions.

In the present model, the compact electron–proton configuration may be viewed as a relativistic standing electron wave surrounding the proton. For the small-hydrogen ground state identified in this paper (Table 3), the electron de Broglie wavelength is $\lambda \approx 17.9$ fm, corresponding to a characteristic oscillation frequency $f_e \approx c/L \approx 1.7 \times 10^{22}$ Hz, where $L$ denotes the effective orbital path length (of order the circumference $L = 2\pi r$). For comparison, the corresponding frequency for ordinary hydrogen is about $7 \times 10^{15}$ Hz.

This motivates the question of whether collective plasma oscillations in extreme astrophysical environments could transiently approach the characteristic frequency scale of small hydrogen. The electron plasma frequency is given by $f_e \approx (e^2 n_e / \varepsilon_0\, m_e)^{1/2} / 2\pi$, where $e$ is the electric charge, $n_e$ is electron density, $\varepsilon_0$ permittivity of vacuum and $m_e$ is the electron mass [30], which reaches values of order $10^{22}$ Hz only for extreme electron densities $n_e \approx 10^{35}$–$10^{36}$ cm$^{-3}$. In collapsing cores, the electron gas is highly degenerate and strongly coupled; the use of the simple plasma-frequency expression here is therefore only an order-of-magnitude indicator of the density scale where collective frequencies enter the $10^{22}$ Hz range.

Such densities are unattainable in laboratory plasmas, but may occur transiently during the collapse of massive stars prior to neutron-star formation. During core collapse, densities increase from about $10^{11}$ g/cm$^3$ (where nuclei begin to dissolve) up to about $10^{14}$ g/cm$^3$ (neutron-star formation). In this intermediate regime, electrons remain relativistic but are not yet universally captured by inverse beta decay ($p + e^- \rightarrow n + \nu_e$), which requires about 0.782 MeV electron energy.

It is therefore to note that brief windows during stellar collapse may reach plasma-frequency scales comparable to those inferred here. Whether this has any connection to actual bound-state formation requires a dedicated dynamical treatment and is not addressed here.

This discussion is intended solely as a plausibility argument identifying environments where the characteristic frequency scales of the model may naturally arise; it does not constitute a formation proof.

Figure A3 illustrates the relationship between plasma frequency and density during stellar collapse. Figure A3 is given just to indicate order-of-magnitude correspondence and does not imply a quantitative formation model.

If small hydrogen were produced cumulatively in energetic stellar environments over cosmic time, its contribution to galactic mass may increase toward low redshift. That is, the trend reported in Ref. [20]—indicating higher baryon dominance in younger galaxies—would be qualitatively consistent with a late-time production scenario, though it does not constitute evidence for it.

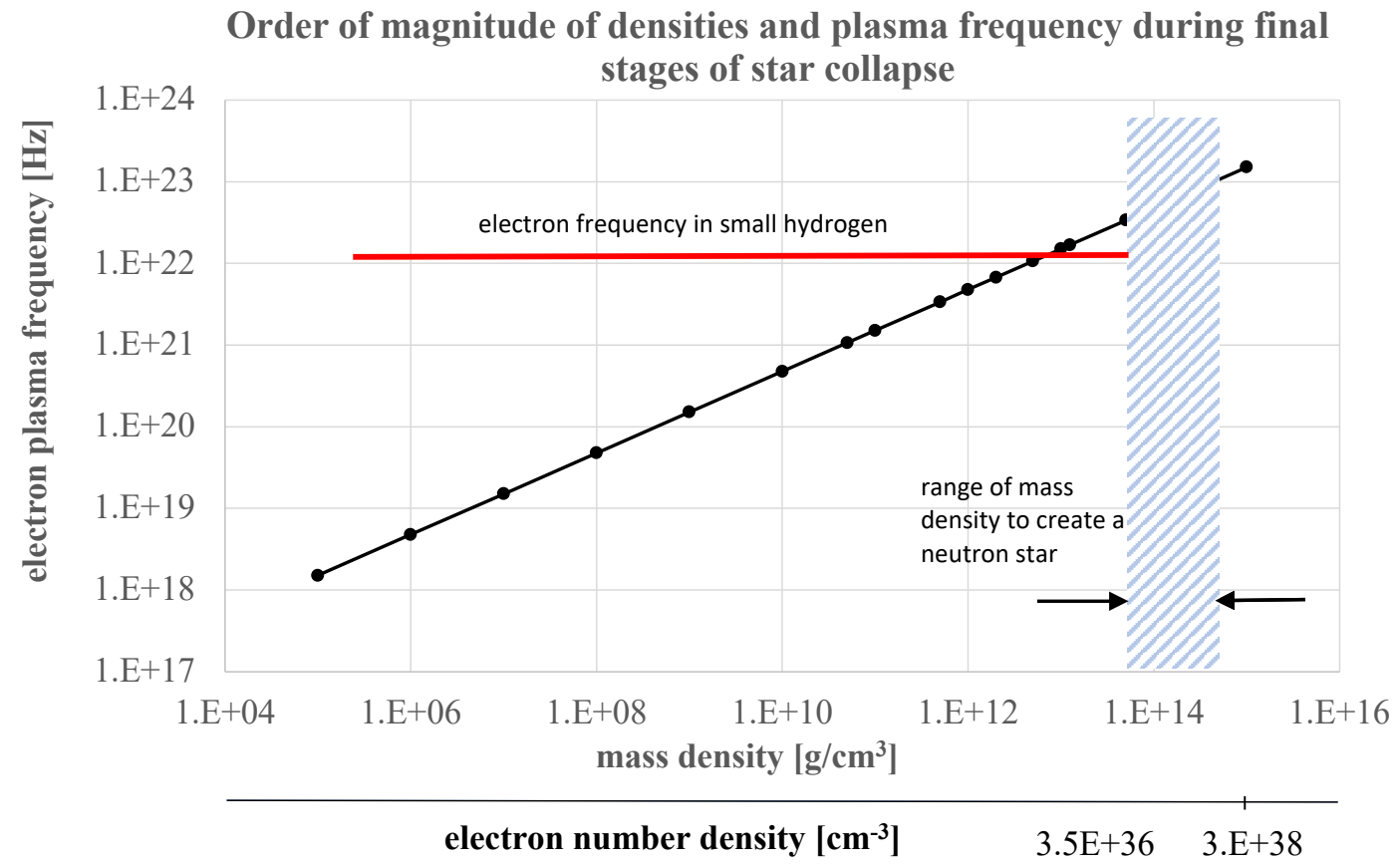

**Figure A3.** Plasma frequency as a function of mass and electron density during the collapse of a large star. The plot indicates the density region where neutron stars form, with slightly lower densities potentially allowing for conditions relevant to the frequency scales discussed here. Dashed lines indicate order-of-magnitude ranges and are drawn to guide the eye,

In addition to core-collapse environments, other relativistic astrophysical plasmas—such as those associated with accretion onto compact objects—may actually provide conditions in which relativistic e–p systems persist over extended timescales. No attempt is made here to estimate formation efficiencies in such environments.

Before discussing observational constraints, it is worth to emphasize that the following comparison is not intended as evidence for compact e–p configurations, but just an illustration of which spectral domains to constrain the phenomenological energy scales discussed in this paper. The characteristic energy scales suggested by the ansatz lie in the range $E_0 \approx 200$–310 keV, with a spin-dependent scale near approximately 100 keV and a central scale near about 260 keV (Sections 2–4).

These scales can be constrained in two observationally distinct regimes. (a) Emission generated in nearby or low-redshift environments ($z \ll 1$) would appear close to the intrinsic energy scale and can actually produce comparatively narrow features in the 200–310 keV band — if radiative channels exist. (b) Emission produced over quite a broad range of high redshifts would be redshifted according to $E_{\text{Observed}} = E_0/(1 + z)$, mapping $z \approx 5$–25 into the present-day approximately 10–50 keV range. Even intrinsically narrow rest-frame features would then be quite s broadened by redshift smearing, producing a smooth excess rather than a sharp line.

The observed smoothness of the extragalactic $X$-ray/$\gamma$-ray background therefore provides strong upper limits on any non-standard high-redshift keV–MeV emission mechanism.

Figure A4 shows the extragalactic $X$-ray and $\gamma$-ray background across both regimes. The gray band highlights the 200–310 keV region relevant for potential low-redshift line searches, while the broad peak near 10–50 keV corresponds to energies where redshifted high-$z$ emission believed to appear. The observed smoothness of the spectrum in both regions therefore places strong constraints on any non-standard keV–MeV emission mechanism.

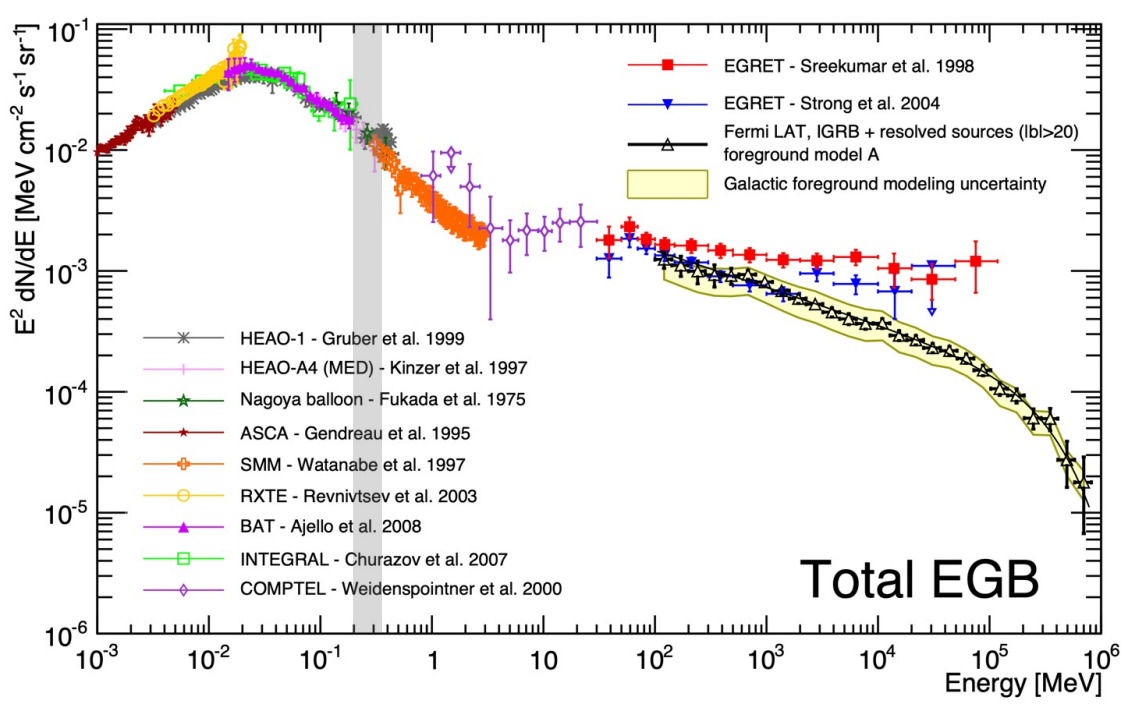

**Figure A4.** Extragalactic $X$-ray and $\gamma$-ray background (EGB) spectrum from various measurements [31]. The gray band indicates the 200–310 keV energy range where relatively nearby sources of compact electron–proton states could, in principle, produce proposed narrow spectral features.

Highly magnetized neutron stars (magnetars) represent another extreme plasma environment that could be examined in future searches for weak or narrow spectral features in the 200–310 keV band, although no such evidence is claimed here.

## References

1. Reeves, R. *A Force of Nature: The Frontier Genius of Ernest Rutherford*; Atlas Books, LLC; W.W. Norton & Company, Inc.: New York, NY, USA, 2008; p. 114. Available online: https://archive.org/details/forceofnaturefro0000reev_o1y4 (accessed on 13 February 2026).
2. Pais, A. *Inward Bound: Of Matter and Sources in the Physical World*; Clarendon Press/Oxford Universty Press: New York, NY, USA, 1986; p. 401. Available online: https://archive.org/details/inwardboundofmat00pais_0 (accessed on 13 February 2026).
3. Schiff, L.I. *Quantum Mechanics*; McGraw–Hill, Inc.:: New York, NY, USA, 1968; p. 470. Available online: https://ia601609.us.archive.org/11/items/ost-physics-schiff-quantummechanics/Schiff-QuantumMechanics.pdf (accessed on 13 February 2026).
4. Maly, J.A.; Va'vra, J. Electron transitions on deep Dirac levels I. *Fusion Sci. Technol*. **1993,** *24*, 307–318. https://doi.org/10.13182/FST93-A30206
5. Maly, J.A.; Va'vra, J. Electron transitions on deep Dirac levels II. *Fusion Sci. Technol*. **1993,** *27*, 59–70. https://doi.org/10.13182/FST95-A30350
6. Smith, F.C.; Johnson, W.R. Relativistic Self-Consistent Fields with Exchange. *Phys. Rev.* **1967**, *160*, 136–142.
7. Bush, B.W.; Nix, J.R. Classical hadrodynamics: Foundations of the theory. *Ann. Phys.* **1993**, *227*, 97–150.
8. Salpeter, E.E.; Bethe, H. A relativistic equation for bound-state problems. *Phys. Rev.* **1951**, *84*, 1232–1242.
9. Spence, J.R.; Vary, J.P. Electron-proton resonances at low energy from a relativistic two-body wave equation. *Phys. Lett. B* **1991**, *271*, 27–31.
10. Brodsky, S.J.; Pauli, H.-C.; Pinsky, S.S. Quantum chromodynamics and other field theories on the light cone. *Phys. Rep*. **1998**, *301*, 299–486.
11. Sommerfeld, A. Zur Quantentheorie der Spektrallinien. *Ann. d. Phys.* **1916**, *356*, 1–94. https://doi.org/10.1002/andp.19163561702

12. Dirac, P.A.M. The quantum theory of the electron. *Proc. R. Soc. Lond. A* **1928**, *117*, 610–624. https://doi.org/10.1098/rspa.1928.0023
13. Flűgge, S. *Practical Quantum Mechanics*; Springer: Berlin/Heidelberg, Germany, 1999. https://doi.org/10.1007/978-3-642-61995-3
14. Adamenko, S.V.; Vysotskii, V.I. Mechanism of synthesis of superheavy nuclei via the process of controlled electron-nuclear collapse. *Found. Phys. Lett.* **2004**, *17*, 203–233. https://doi.org/10.1023/B:FOPL.0000032473.74790.38
15. Paillet, J.L.; Meulenberg, A. Advance on Electron Deep Orbits of the Hydrogen Atom. *J. Condens. Matter Nucl. Sci.* **2017**, *24*, 258–277.
16. Gaite, J. The relativistic virial theorem and scale invariance. *Phys.-Usp.* **2013**, *56*, 919–931. https://doi.org/10.3367/UFNe.0183.201309f.0973
17. Lucha, W. Relativistic virial theorems. *Mod. Phys. Lett. A* **1990**, *5*, 2473–2483.
18. Goldstein, H. *Classical Mechanics*. Addison–Wesley Publishing Company, Inc.: Reading, MA, 1980. Available online: https://www.scribd.com/document/396594738/Addison-Wesley-Series-in-Physics-Herbert-Goldstein-Classical-Mechanics-Addison-Wesley-Pub-Co-1980 (accessed 13 February 2026).
19. Hwang, D.S.; Kim, C.S.; Namgung, W. Average kinetic energy of heavy quark and virial theorem. *Phys. Lett. B* **1997**, *406*, 117–122.
20. Genzel, R.; Förster Schreiber, N.M.; Übler, H.; Lang, P.; Naab, T.; Bender, R.; Tacconi, L.J.; Wisnioski, E.; Wuyts, S.; Alexander, T.; et al. Strongly baryon-dominated disk galaxies at the peak of galaxy formation ten billion years ago. *Nature* **2017**, *543*, 397–401. https://doi.org/10.1038/nature21685.
21. Comerón, S.; Trujillo, I.; Cappellari, M.; Buitrago, F.; Garduño, L.E.; Zaragoza-Cardiel, J.; Zinchenko, I.A.; Lara-López, M.A.; Ferré-Mateu, A.; Dib, S. The massive relic galaxy NGC 1277 is dark matter deficient. *Astron. Astrophys.* **2023**, *675*, A143.
22. Wagoner, R. Big-bang nucleosynthesis revisited. *Astrophys. J.* **1973**, *179*, 343–360.
23. Markevitch, M.; Gonzalez, A.H.; Clowe, D.; Vikhlinin, A.; Forman, W.; Jones, C.; Murray, S.; Tucker, W. Direct constraints on the dark matter self-interaction cross-section from the merging galaxy cluster 1E 0657-56. *Astrophys. J.* **2004**, *606*, 819–824.
24. Harvey, D; Massey, R.; Kitching, T.; Taylor, A.; Tittley, E. The non-gravitational interactions of dark matter in colliding galaxy clusters. *Science*, **2015**, *347*, 1462–1465. https://doi.org/10.1126/science.1261381
25. Weidenspointner, G.; Kiener, J.; Gros, M.; Jean, P.; Teegarden, B.J.; Wunderer, C.; Reedy, R.C.; Attié, D.; Diehl, R.; Ferguson, C.; et al. First identification and modelling of SPI background lines. *Astron. Astrophys.* **2003**, *411*, L113–L116.
26. Schott, G.A. The electromagnetic field of a moving uniformly and rigidly electrified sphere and its radiationless orbits. *Philos. Mag.* **1933**, *15*, 752–761. https://doi.org/10.1080/14786443309462219
27. Goedecke, G.H. Classically radiation-less motions and possible implications for quantum theory. *Phys. Rev.* **1964**, *135*, B281–B288.
28. Okninski, A. A Remark on Zitterbewegung and Schott's Radiationless Motion. August 2021. Available online: https://www.researchgate.net/publication/354117700 (accessed on 13 February 2026).
29. Barut, A.O.; Bracken, A.J. Zitterbewegung and the internal geometry of the electron. *Phys. Rev. D* **1981**, *23*, 2454–2463.
30. Gibbon, P. Introduction to plasma physics. In *Plasma Wake Acceleration (CERN-2016-001)*; Holzer, B. (Ed.); CERN: Geneva, Switzerland, 2016. ; pp. 51–66. http://dx.doi.org/10.5170/CERN-2016-001
31. Ackermann, M.; Ajello, M.; Albert, A.; Atwood, W.B.; Baldini, L.; Ballet, J.; Barbiellini, G.; Bastieri, D.; Bechtol, K.; Bellazzini, R.; et al. The spectrum of isotropic diffuse gamma-ray emission between 100 MeV and 820 GeV. *Astrophys. J.* **2015**, *799*, 86.